\documentclass[preprint]{aastex}

\newcommand{\objC}{J1632--0033}

\slugcomment{}

\begin{document}

\title{ PMN~\objC: A new gravitationally lensed quasar }

\author{
Joshua N.\ Winn\altaffilmark{1,2,3},
Nicholas D.\ Morgan\altaffilmark{1},
Jacqueline N.\ Hewitt\altaffilmark{1},
Christopher S.\ Kochanek\altaffilmark{4},
James E.J.\ Lovell\altaffilmark{5},
Alok R.\ Patnaik\altaffilmark{6},
Bart Pindor\altaffilmark{7},
Paul L.\ Schechter\altaffilmark{1,2},
Robert A.\ Schommer\altaffilmark{8}
}

\altaffiltext{1}{Department of Physics, Massachusetts Institute of Technology,
    Cambridge, MA 02139}
\altaffiltext{2}{Visiting Astronomer, Cerro Tololo Inter-American Observatory,
    National Optical Astronomy Observatories}
\altaffiltext{3}{NSF Astronomy \& Astrophysics Postdoctoral Fellow,
    Harvard-Smithsonian Center for Astrophysics,
    60 Garden St., Cambridge, MA 02138}
\altaffiltext{4}{Harvard-Smithsonian Center for Astrophysics,
    60 Garden St., Cambridge, MA 02138}
\altaffiltext{5}{Australia Telescope National Facility, CSIRO, PO Box 76,
    Epping, NSW 1710, Australia}
\altaffiltext{6}{Max-Planck-Institut f\"{u}r Radioastronomie,
    Auf dem H\"{u}gel 69, 53121 Bonn, Germany}
\altaffiltext{7}{Princeton University Observatory, Peyton Hall,
    Princeton, NJ 08544-1001}
\altaffiltext{8}{Cerro Tololo Inter-American Observatory, National Optical
    Astronomy Observatories, Casilla 603, La Serena, Chile}

\begin{abstract}
We report the discovery of a gravitationally lensed quasar resulting
from our survey for lenses in the southern sky.  Radio images of
\objC~with the VLA and ATCA exhibit two compact, flat-spectrum
components with separation $1\farcs47$ and flux density ratio 13.2.
Images with the HST reveal the optical counterparts to the radio
components and also the lens galaxy.  An optical spectrum of the
bright component, obtained with the first Magellan telescope, reveals
quasar emission lines at redshift 3.42.  Deeper radio images with
MERLIN and the VLBA reveal a faint third radio component located near
the center of the lens galaxy, which is either a third image of the
background quasar or faint emission from the lens galaxy.
\end{abstract}

\keywords{gravitational lensing, quasars: individual (\objC)}

\section{Introduction}
\label{sec:intro}

When a galaxy happens to lie along nearly the same line of sight as a
more distant object, gravitational lensing may cause multiple images
of the background object to appear in the sky. These cases of
``strong'' lensing (as opposed to ``weak'' lensing, in which images of
background objects are distorted but not multiply imaged) can be used
to measure the masses (see, e.g., \citealt{kochanek95}), evolution
(\citealt{kochanek00}) and extinction laws (\citealt{falco99}) of
distant galaxies, and to determine the Hubble constant
\citep{refsdal64,koopmans99} and cosmological constant
\citep{turner90,fukugita90,falco98}, among other applications
\citep[for reviews, see][]{narayan98,blandford92}.

We have been conducting a survey for new lenses to be used in these
applications. Ours is the latest in a series of VLA-based surveys for
gravitational lenses \citep{mgvla,jvas,browne00}, and is the first to
explore the portion of the southern sky accessible to the VLA
($0\arcdeg > \delta > -40\arcdeg$).

The search strategy relies on the fact that the great majority ($\sim
95\%$) of flat-spectrum radio sources appear as point sources in
8.5~GHz VLA A-array images (which have $\sim 0\farcs3$ resolution).
Our initial sample of 4097 flat-spectrum radio sources was generated
by cross-correlating the PMN (4.85~GHz; \citealt{pmn}) and NVSS
(1.4~GHz; \citealt{nvss}) catalogs.  Those few sources exhibiting
multiple compact components in 8.5~GHz VLA images, with separations
ranging from $0\farcs 3$ to $6\arcsec$, were selected as lens
candidates. To filter out core-jet sources, the candidates underwent
radio imaging with higher angular resolution. The very best candidates
were then observed with optical telescopes, in order to search for a
lens galaxy.

The object presented in this paper, \objC, is the fourth confirmed
lens to date.  Section~\ref{sec:radio} presents the radio
observations.  Section~\ref{sec:optical} presents optical data, from
Hubble Space Telescope imaging (\S~\ref{subsec:hst}), ground-based
direct imaging (\S~\ref{subsec:ground}) and spectroscopy
(\S~\ref{subsec:spectrum}).  Simple gravitational lens models are
discussed in \S~\ref{sec:models}.  The nature of the third faint radio
component is also taken up in this section, along with possible
implications for the mass distribution of the lens galaxy if the third
component is an additional quasar image.  Finally,
\S~\ref{sec:summary} summarizes the evidence that \objC~is a
gravitationally lensed quasar, and describes prospects for future
observations that would use the system as a probe of cosmology or
galactic structure.

\section{Radio observations---overview}
\label{sec:radio}

Table~\ref{tbl:radio} lists the radio observations with the VLA,
VLBA\footnote{The Very Large Array (VLA) and Very Long Baseline Array
(VLBA) are operated by the National Radio Astronomy Observatory, a
facility of the National Science Foundation operated under cooperative
agreement by Associated Universities, Inc.}, MERLIN\footnote{The
Multi-Element Radio Linked Interferometry Network (MERLIN) is a UK
national facility operated by the University of Manchester on behalf
of SERC.}, and ATCA\footnote{The Australia Telescope Compact Array
(ATCA) is part of the Australia Telescope which is funded by the
Commonwealth of Australia for operation as a National Facility managed
by CSIRO.} in chronological order. The first 3 entries represent
archival VLA data that we obtained and reduced; the remaining 10 are
observations that we conducted after identifying \objC~as a
gravitational lens candidate.

In all the images except the VLBA image and the second (deeper) MERLIN
image, there are two components separated by $1\farcs47$.  We refer to
the northwest (brighter) component as A, and the southeast (dimmer)
component as B.

The deeper MERLIN image revealed a faint third component at the
$6.7\sigma$ level (where $\sigma$ is the RMS noise in a nearby blank
area of the radio image).  It is located close to B and nearly along
the line between A and B.  This component is also present at the
$4.5\sigma$ level in the VLBA image, although we noticed it only after
seeing it in the MERLIN image.  We refer to this faint component as C.
Component C was not detected in any of the other images, nor would it
be expected to be detectable in those images, given their lower
angular resolution and/or higher noise level.

\begin{figure}
\figurenum{1}
\plotone{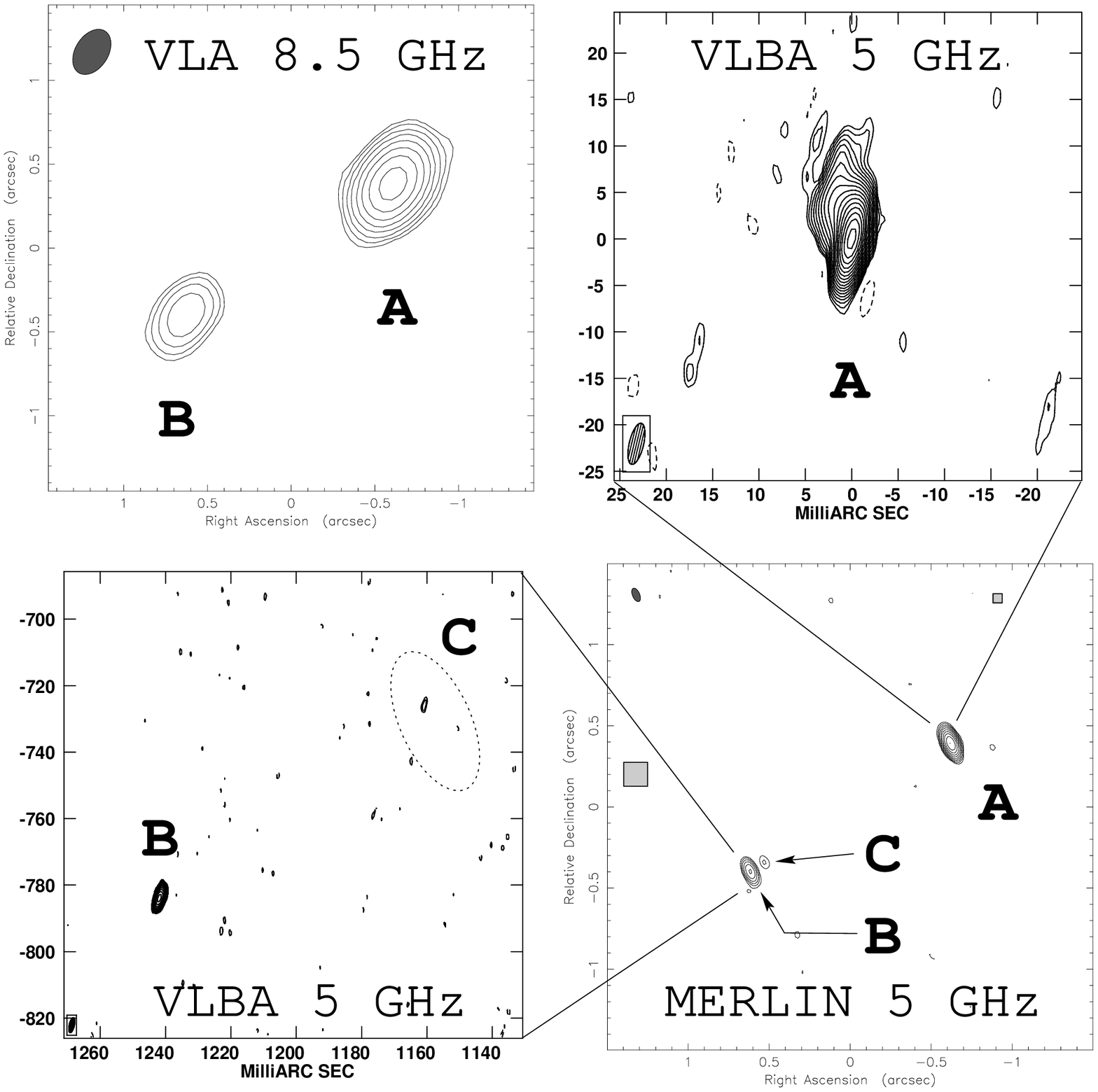}
\caption{ \scriptsize Radio contour maps of \objC. In each panel, the
synthesized beam is illustrated in one corner.  {\bf Upper left.}
8.5~GHz VLA observation of 1994~February~24.  Contours begin at
$3\sigma$ and increase by factors of two, where
$\sigma=0.26$~mJy~beam$^{-1}$.  {\bf Lower right.} 5~GHz MERLIN
observation of 2001~April~13.  Contours begin at $3\sigma$ and
increase by factors of two, where $\sigma=0.13$~mJy~beam$^{-1}$.  The
small gray squares illustrate the dimensions of the VLBA subimages
centered on component A (upper right panel) and B+C (lower left
panel).  {\bf Upper right.}  From the 5~GHz VLBA observation of
2000~April~29.  Contours begin at $3\sigma$ and increase by factors of
$\sqrt{2}$, where $\sigma=0.13$~mJy~beam$^{-1}$.  {\bf Lower left.}
From the 5~GHz VLBA observation of 2000~April~29.  Contours begin at
$3\sigma$ and increase by factors of $\sqrt{2}$, where
$\sigma=0.11$~mJy~beam$^{-1}$.  The dotted ellipse is centered on the
MERLIN position of component C, and its dimensions are half the
dimensions of the MERLIN synthesized beam.  \normalsize}

\label{fig:radio}
\end{figure}

Components A and B are unresolved in all the images except the VLBA
image, in which component A is partially resolved.  Component C is too
faint to determine whether it is unresolved or slightly resolved with
much confidence.  Figure~\ref{fig:radio} displays a representative VLA
image of the system, the deeper MERLIN image, and higher-resolution
VLBA images of components A and B.  Table~\ref{tbl:radio-positions}
gives the relative positions of A, B, and C, as determined from the
VLBA image.  The J2000 coordinates of \objC, as derived from the VLA
observations, are also given in Table~\ref{tbl:radio-positions}.

\subsection{Data reduction and analysis}

The details of the data reduction and analysis were as follows:

{\bf VLA:} The VLA data were calibrated with standard routines in the
software package AIPS. For the first three VLA observations
(1984--1999), radio source 3C286 was used to set the absolute flux
density scale, using the procedures recommended in the VLA Calibrator
Manual. For the VLA observations in 2000, the source J2355+4950 was
used instead; this compact source is monitored monthly by G.~Taylor
and S.~Myers of NRAO, and has been found to have a stable flux
density. The assumed flux densities of J2355+4950 were 2.306~Jy
(1.4~GHz), 0.602~Jy (15~GHz), 0.473~Jy (22.5~GHz), and 0.284~Jy
(43~GHz). We applied gain-elevation corrections for data at 15~GHz and
higher frequencies based on gain curves prepared by NRAO staff.

Imaging and analysis were performed with the software package Difmap
\citep{difmap}. We fitted a surface-brightness model consisting of 2
unresolved points to the visibility function, and used this model to
perform phase-only self-calibration with a solution interval of 30
seconds. This process, model-fitting and self-calibration, was
repeated (typically 3 times) until the model converged. For the
1.4~GHz data, which had the lowest spatial resolution, we fixed the
relative separation and orientation of the model components at the
values obtained from the VLBA data.

{\bf ATCA:} The ATCA data were calibrated using the software package
MIRIAD \citep{miriad}. The absolute flux density scale of the ATCA
data was set by observations of PKS~B1934--638. The imaging and
analysis steps were the same as for the VLA data. As with the 1.4~GHz
VLA data, the relative separation of the two model components was kept
fixed at the values derived from the VLBA data, in order to improve
the accuracy with which we could extract separate flux densities for A
and B.

{\bf MERLIN:} Calibration of the MERLIN data was performed at Jodrell
Bank using both standard MERLIN software and AIPS. The absolute flux
scale was set by observing 3C286 and assuming a flux density of
7.38~Jy on the shortest baseline. Imaging and analysis were performed
with Difmap in the same manner as for the VLA and ATCA data.  For the
second, much longer MERLIN observation, we also performed one round of
amplitude self-calibration, with a solution interval of 30 minutes.
In that case the residuals (after the two-component model was
subtracted) indicated the presence of a third component.  Our final
model consisted of three point sources (A, B, and C).

{\bf VLBA:} For the VLBA observation, the total observing bandwidth
was divided into 4 intermediate frequency bands, each of which was
subdivided into $16\times0.5$~MHz channels. Fringe-fitting,
calibration, and imaging of the VLBA data were all performed with
AIPS.  After fringe-fitting, we reduced the data volume by averaging
in time into 6-second bins and in frequency into 1~MHz bins. These
values were chosen to reduce the data volume as much as possible while
keeping the amount of bandwidth smearing and time-average smearing
below $1\%$ over the required field of view.

Components A and B were obvious in the ``dirty'' image (prior to
deconvolution). We employed the multiple-field implementation of the
CLEAN algorithm, with a 180~mas~$\times$~180~mas field centered on
each component. The model developed by the CLEAN algorithm was used to
self-calibrate the antenna phases with a solution interval of 30
seconds. We iterated this process 5 times before arriving at the final
model.  The images based on this model, with uniform weighting and an
elliptical Gaussian restoring beam, are displayed in
Figure~\ref{fig:radio}.

After seeing component C in the MERLIN image, we repeated the imaging
step using two 200~mas~$\times$~200~mas fields, one centered on
component A, and the other centered between B and the expected
location of C.  We found a $4.5\sigma$ peak within 5~mas of the MERLIN
position of C.  It is the brightest peak within the field, and does
not belong to the sidelobe pattern of either of the two brighter
components A or B.  We conclude that component C is also present in
the VLBA image.

\subsection{Radio observations---discussion}
\label{subsec:radio-discuss}

\begin{figure}
\figurenum{2}
\plotone{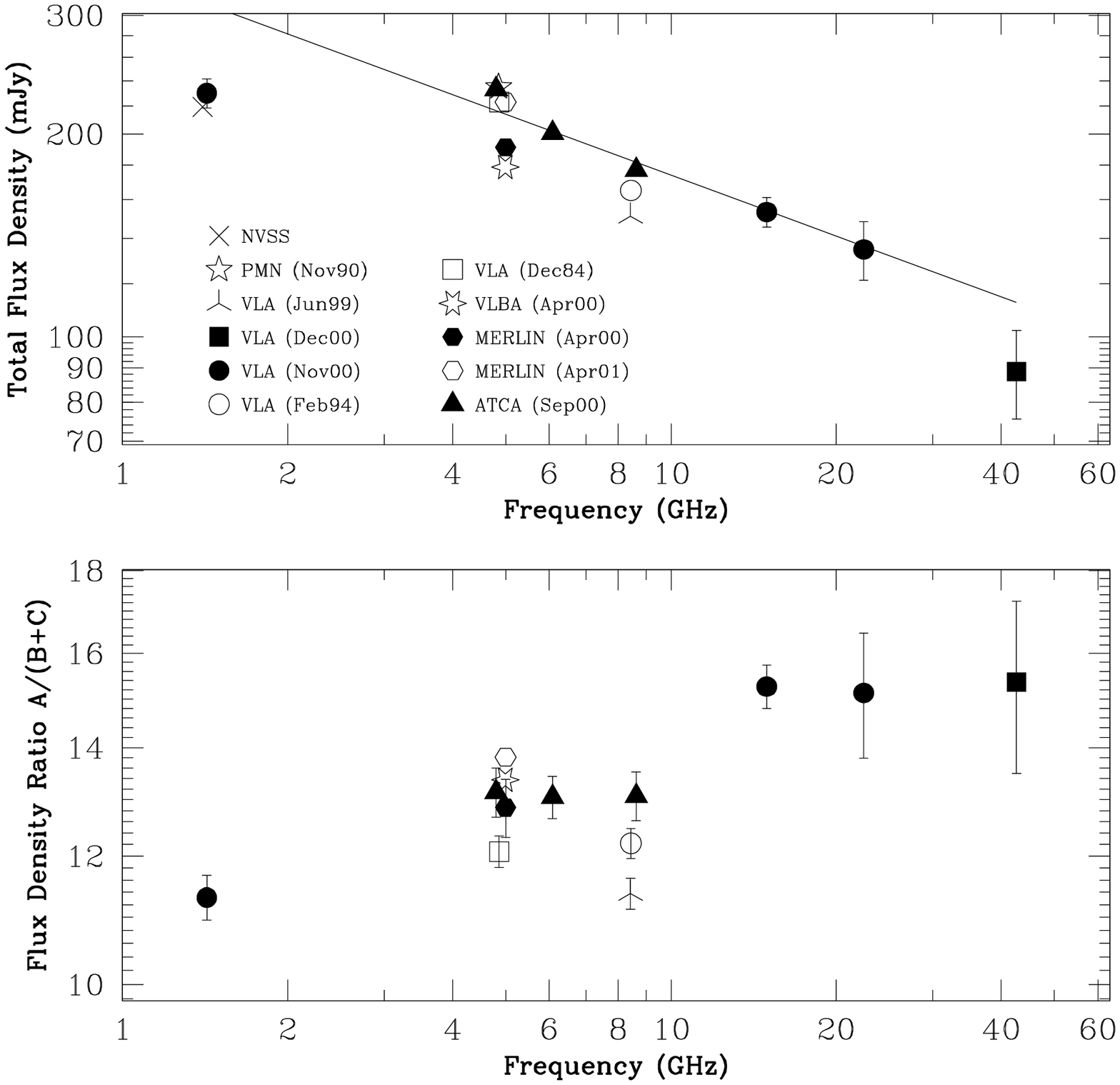}
\caption{ {\bf Top panel.} Total flux density of \objC~as a function
of radio frequency.  For reference, a power law $S_\nu \propto
\nu^{-0.3}$ is also plotted.
\vskip 0.2in {\bf Bottom panel.} Flux density ratio as a function of
radio frequency. The error estimates were computed by assuming that
the uncertainty in the relative flux density scale is equal to the RMS
noise level in the image. }
\label{fig:fluxes}
\end{figure}

The flux densities of A, B, and C (when detected) are reported in
Table~\ref{tbl:radio} for each observation, along with the RMS noise
level and an estimate of the uncertainty in the absolute flux
scale. These numbers were used to generate the two plots presented in
Figure~\ref{fig:fluxes}.

The top panel of Figure~\ref{fig:fluxes} is a logarithmic plot of the
total flux density of both components as a function of radio
frequency.  Evidently \objC~has a fairly flat radio spectrum, typical
of radio-loud quasars (and indeed, the optical spectrum presented in
\S~\ref{subsec:spectrum} verifies that it is a quasar).  From 5~GHz to
22.5~GHz, the data that were all taken in the year 2000 are a good
match to the power law $S_\nu \propto \nu^{-0.3}$.  At 8.5~GHz, there
is evidence for variability at the 5--10\% level on a time scale of
years. (The discrepancies between the 5~GHz measurements are more
difficult to interpret, because the VLBA and MERLIN probe much smaller
angular scales than the other 5~GHz observations.)

Two compact, flat-spectrum radio components separated by $1\farcs47$
are likely to be either a pair of quasars at the same redshift (a
binary quasar) or a pair of gravitationally lensed images of a single
quasar. In fact, \objC~is a lens rather than a binary quasar.  The
most definitive evidence, optical detection of the lens galaxy, is
presented in the next section.  But, interestingly, lensing is
implicated by the radio evidence alone.

The first clue is the near-equality of spectral indices of components
A and B.  This is a natural consequence of gravitational lensing,
which is achromatic, but would require a coincidence under the binary
quasar hypothesis.  The bottom panel of Figure~\ref{fig:fluxes} is a
logarithmic plot of the A/B flux density ratio as a function of
frequency.  The ratio increases by less than 30\% over a factor of 30
in frequency (and even this may be due in part to components B and C
being lumped together in the low-resolution images---see
\S~\ref{subsec:third}).  Assuming that component A obeys $S_{\nu}(A)
\propto \nu^{\alpha}$, and likewise $S_{\nu}(B) \propto \nu^{\beta}$,
then a least-squares analysis implies $\alpha - \beta = 0.07\pm 0.01$.
We estimate the probability of randomly drawing two quasars with
spectral indices that match as closely to be 7\%, by analyzing the
histogram of spectral indices of flat-spectrum sources in the PMN
tropical and equatorial catalogs.

The second piece of radio evidence, which is more compelling, is the
location of the faint component C.  It is located almost exactly where
one would expect to find the center of a lens galaxy---along the line
joining A and B, and closer to the dimmer of the two components, as
simple lens models predict (see \S~\ref{subsec:models}).  Regardless
of whether C is emission from the lens galaxy or a third, demagnified
image of the background quasar (see \S~\ref{subsec:third}), the
existence and location of C implies that the system is a lens rather
than a binary quasar.

\section{Optical observations}
\label{sec:optical}

\subsection{HST images}
\label{subsec:hst}

On 2001~July~1 we obtained optical images of \objC~with the WFPC2
camera aboard the HST\footnote{ Data from the NASA/ESA Hubble Space
Telescope (HST) were obtained from the Space Telescope Science
Institute, which is operated by AURA, Inc., under NASA contract
NAS~5-26555.}.  We obtained 3 dithered exposures through each of two
filters, F555W ($\approx V$) and F814W ($\approx I$).  The total
exposure time for each filter was 2000~seconds.  The target was
centered in the PC chip.

\begin{figure}
\figurenum{3}
\plotone{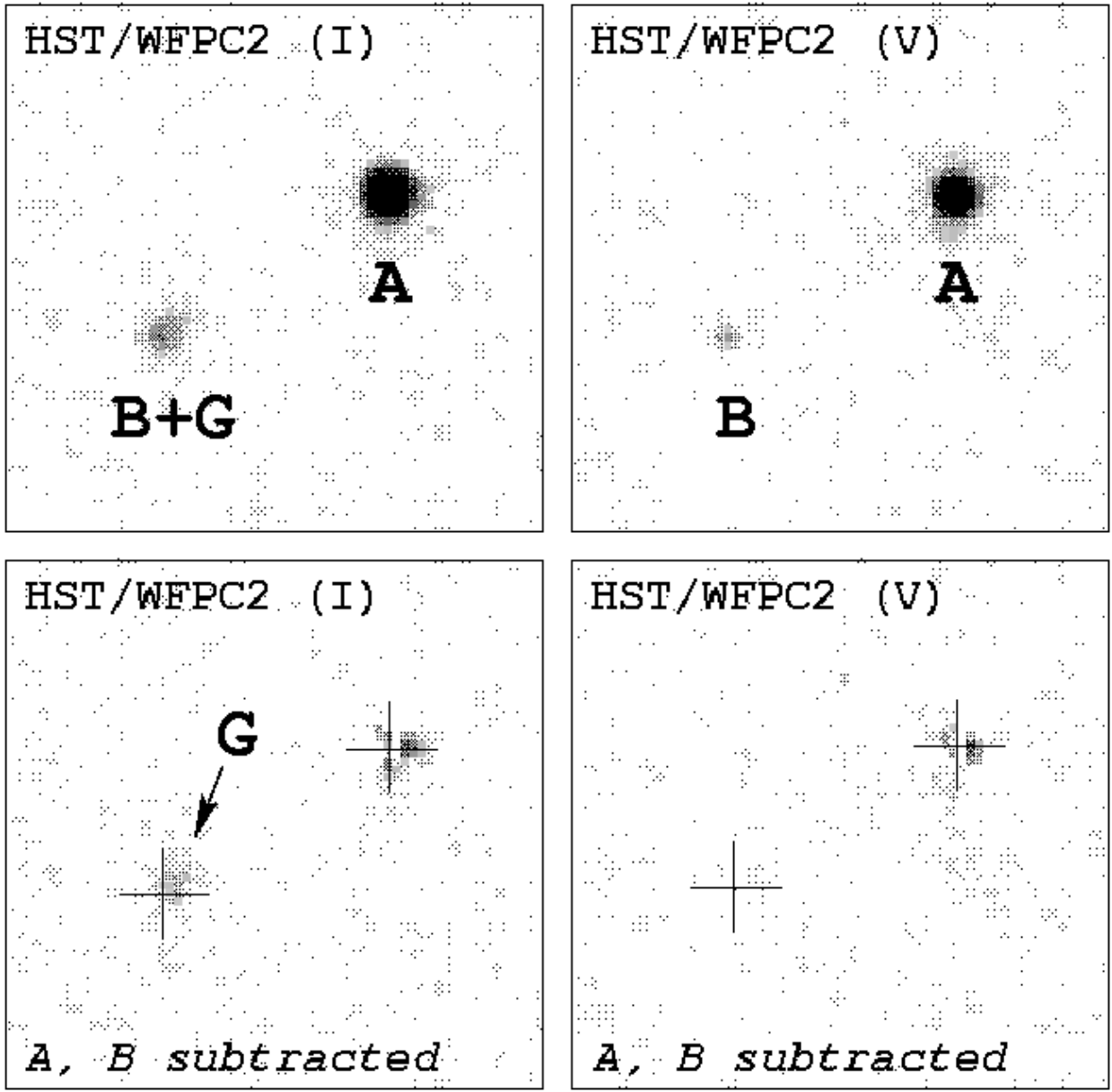}
\caption{ {\bf Top left panel.} Optical image with HST/WFPC2 (F814W,
$\approx I$ band).  {\bf Bottom left panel.}  Same, after subtraction
of a surface-brightness model consisting of two point sources (see
\S~\ref{subsec:hst}).  {\bf Top right panel.} Optical image with
HST/WFPC2 (F555W, $\approx V$ band).  {\bf Bottom right panel.}  Same,
after subtraction of the two point sources from our final 3-component
surface-brightness model (see \S~\ref{subsec:hst}).  The de
Vaucouleurs profile has not been subtracted, in order to highlight
the lens galaxy.  }

\label{fig:hst}
\end{figure}

The exposures were combined and cosmic rays rejected using the Drizzle
algorithm as implemented in IRAF\footnote{ The Image Reduction and
Analysis Facility (IRAF) is a software package developed and
distributed by the National Optical Astronomical Observatories, which
is operated by AURA under a cooperative agreement with the National
Science Foundation.}.  In both filters, light from a bright ($R=11$)
binary star $30\arcsec$ to the south of \objC~caused a large gradient
in the background level across the PC chip, which was removed by
fitting (and then subtracting) a second-order polynomial surface to
the apparently empty regions of the image.  The upper two panels of
Figure~\ref{fig:hst} show the final images.

In the $V$-band image, the optical counterparts to radio components A
and B were detected, and are unresolved.  In the $I$-band image, A and
B are also present, but there is a diffuse object near B that we
identify as the lens galaxy.  We constructed a photometric model for
each image using software written by B.\ McLeod (see, e.g.,
\citealt{lehar00}).  This program finds the parameters of a
surface-brightness model (which consists of a user-determined number
of point sources and simple galaxy profiles) convolved with a PSF
computed by the program ``Tiny Tim'' \citep{tinytim}, such that the
mean-square residuals are minimized.

For the $I$-band image, we tried initially to fit a model consisting
of 2 point sources only.  Subtracting two point sources from the
$I$-band image leaves behind residuals near A with peak intensity
$\sim 3\%$ the central intensity of A, which in our experience is the
level produced by inaccuracies in the Tiny Tim PSF.  However, the
residuals near B have peak values 40\% the central intensity of B and
$8\sigma$, where $\sigma$ is the RMS level in a blank region of the
image.  This indicates there is a diffuse source of unmodeled light
near B: the lens galaxy.

Our final model consisted of two points and a circular de Vaucouleurs
profile.  Upon subtraction of this model, the residuals near B are
consistent with random noise.  In order to highlight the galaxy, the
lower right panel of Figure~\ref{fig:hst} shows the image that results
when only the point sources of the best-fit model have been
subtracted.  Table~\ref{tbl:hst} lists the parameters of the best-fit
photometric model.

As mentioned earlier, the position of the center of the galaxy derived
from the HST image is consistent with the location of radio component
C, and both are located where simple lens models predict that a lens
galaxy (or demagnified third image) to appear.  However, because it is
not clear whether the galaxy itself is responsible for the radio
emission, we have labeled the optical galaxy G rather than C.

For the $V$-band image, our model consisted of two point sources.  The
residuals (after the best-fit model was subtracted from the original
image) are shown in the lower left panel.  The peak residuals near A
are 3--5\% of the central intensity of A.  The peak residuals near B
are dominated by noise rather than PSF subtraction; they are $3\sigma$
where $\sigma$ is the RMS level in a blank region of the image.  The
separation and position angle of the points are consistent with the
A/B radio values.

For the final $V$-band results listed in Table~\ref{tbl:hst}, we also
added a de Vaucouleurs profile to the model, but we fixed its
effective radius and fixed all the positions of the model components
at the $I$-band values, and allowed only their fluxes to vary.
Because the $V$-band image does not provide independent evidence for
the galaxy, the $V$-band galaxy magnitude should probably be
considered as an approximate lower limit.

The zero-points of the magnitude scales were 21.69 for $I$ and 22.54
for $V$, as determined by \citet{dolphin00} and corrected for infinite
aperture.  We also applied a correction for losses due to the
charge-transfer-efficiency problem of WFPC2, as per the prescription
of \citet{dolphin00}.  These corrections were large: for component A
they were $+0.20$~mag for both filters, and for components B+G they
were $+0.50$ for $I$ and $+0.90$ for $V$.  The quoted magnitudes are
therefore considerably uncertain and should be treated with caution.

\subsection{Ground-based images}
\label{subsec:ground}

Before obtaining the HST image we attempted to search for evidence of
a lens galaxy in ground-based optical images.  Although the HST image
contains the most convincing evidence of the lens galaxy, the
ground-based results are useful in providing total magnitudes and flux
ratios over a wider range of wavelengths.

On 2000~July~25 we obtained $BVRI$ optical images of \objC~with the
Mosaic II camera on the prime focus of the Blanco 4-meter telescope at
CTIO\footnote{ Cerro Tololo Inter-American Observatory (CTIO) is
operated by the Association of Universities for Research in Astronomy
Inc., under a cooperative agreement with the National Science
Foundation as part of the National Optical Astronomy Observatories.}.
The night was photometric. Each exposure lasted 600 seconds.  After
extracting the images from chip \#2, we corrected them for cross-talk
from the paired amplifier, bias-subtracted and flat-fielded them with
standard IRAF procedures, and defringed the $I$-band images using a
fringe template kindly supplied by R.C.\ Dohm-Palmer.

The seeing was $1\farcs2$.  The object could barely be resolved in the
$I$-band image but appeared pointlike in the other images.  A
synthetic circular aperture of radius $7\arcsec$ was used to compute
total instrumental magnitudes (after subtracting all neighboring
objects within $14\arcsec$).  To place the instrumental magnitudes on
a standard photometric system, we also observed the standard stars
\#355, \#360, and \#361 from field SA110 described by
\citet{landolt92}.  Our photometric solutions took the form:
\begin{equation}
m_{\mbox{std}} = m_{\mbox{inst}} + c_1 + c_2 (B_{\mbox{inst}}-R_{\mbox{inst}}),
\end{equation}
where
\begin{equation}
m_{\mbox{inst}} = -2.5\log\left(\frac{\mbox{counts}}{\mbox{time}}\right) - k_m\times{\mbox{airmass}}.
\end{equation}
We adopted ``typical'' CTIO coefficients of $k_I = 0.06$, $k_R =
0.11$, $k_V = 0.15$, and $k_B = 0.28$ \citep{landolt92}. Star \#355
was not used for the $I$-band solution because it was overexposed.
Table~\ref{tbl:optical} reports the calibrated magnitudes of \objC.
The $I$-band magnitude disagrees with the F814W magnitude reported in
\S~\ref{subsec:hst}, possibly due to the CTE problem of WFPC2.

On 25--26~March 2001 we obtained optical images of \objC~with MagIC, a
CCD camera at the Nasmyth focus of the 6.5-meter Baade
telescope.\footnote{The 6.5-meter Baade telescope is the first
telescope of the Magellan Project, a collaboration between Carnegie
Observatories, the University of Arizona, Harvard University, the
University of Michigan, and MIT.} We obtained 480-second exposures in
each of the Sloan filters $g'$, $r'$, and $i'$ in $0\farcs8$ seeing.
On 11--12~June 2001 we also obtained an additional three $i'$
exposures totaling 1350 seconds, in $0\farcs6$ seeing.

The optical counterpart of \objC~appeared double in all images.  In
particular, in the $i'$-images from June, the southeast component was
slightly resolved rather than pointlike, as would be expected from a
combination of quasar B and the lens galaxy.

For each filter, we estimated the flux ratio A:(B+G) using the
following procedure. With the DAOPHOT package in IRAF, we constructed
an empirical PSF of radius $3\arcsec$, using the signal-weighted
average of several bright and isolated field stars.  We then fit a
model consisting of two point sources to the optical counterpart of
\objC.  The results for the flux ratio are printed in
Table~\ref{tbl:fluxratio}.

The ratio decreases with wavelength for two reasons.  First, G is
redder than the quasar images, as demonstrated by the analysis of the
HST images in \S~\ref{subsec:hst}.  Second, the reddening of the
quasar light due to passage through the lens galaxy is apparently
larger for image B than for image A.  This follows from the
observation that the flux ratios in the $g'$ and $V$-band images
(which do not contain a significant contribution from lens galaxy
light) are larger than the radio value.  This is reasonable, since B
is considerably closer to the galaxy center than A.

\subsection{Optical spectrum}
\label{subsec:spectrum}

On 2000~September~1, we obtained an optical spectrum of \objC~with the
3.5-meter telescope at Apache Point Observatory\footnote{The Apache
Point Observatory 3.5-meter telescope is owned and operated by the
Astrophysical Research Consortium.}. We used the Double Imaging
Spectrograph in low resolution mode. The processed spectrum had a very
weak signal, with two features at the $1-2\sigma$ level whose
wavelength ratio was consistent with the common quasar emission lines
Ly$\alpha$ (1216\AA) and \ion{C}{4} (1549\AA), allowing the tentative
identification of \objC~as a quasar at redshift 3.42.

This identification was confirmed on 2001~March~23, when we obtained
an optical spectrum with the 6.5-meter Baade telescope.  We used a
Boller \& Chivens slit spectrograph with a 600 lines~mm$^{-1}$
grating, giving a pixel scale of $0\farcs44$~pixel$^{-1}$, dispersion
2.75\AA~pixel$^{-1}$, resolution $10$\AA, and wavelength coverage
4500--7300\AA. We used the WG360 Schott glass blocking filter to block
second order contamination. The slit was $1\farcs3$ wide, centered on
component A, and oriented perpendicular to the A/B position angle.

\begin{figure}
\figurenum{4}
\plotone{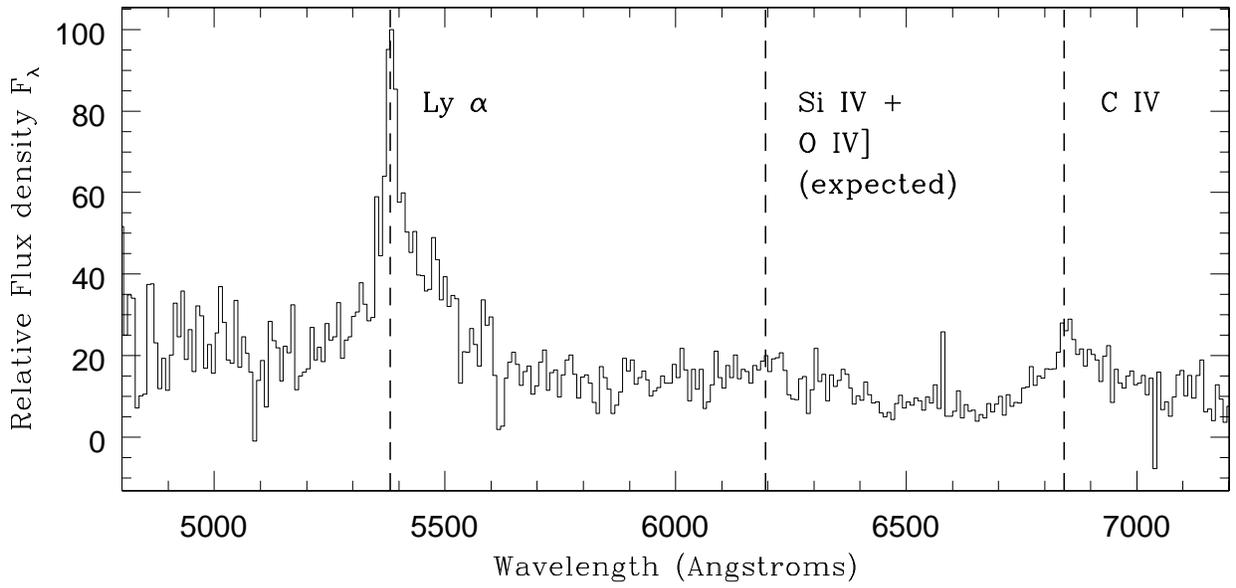}
\caption{ Optical spectrum of component A of \objC~(see
\S~\ref{subsec:spectrum}), with resolution $\sim 10$\AA.
The redshift is $z=3.424\pm 0.007$. }
\label{fig:spectrum}
\end{figure}

Flat-fielding, spectrum extraction, and wavelength calibration were
carried out with standard IRAF procedures. The processed spectrum is
shown in Figure~\ref{fig:spectrum}. Here the Ly$\alpha$ and \ion{C}{4}
emission lines are detected with high significance, implying
$z=3.424\pm 0.007$. The expected position of \ion{Si}{4}~+~\ion{O}{4}]
(1400\AA) is also shown.

As for the lens galaxy, its redshift is currently unknown, but it is
possible to estimate photometrically.  \citet{kochanek00} devised a
method to estimate the redshift of a lens galaxy, given the redshift
of the background source, the image separation, and the magnitudes and
effective radius of the galaxy.  The idea is to require that the
galaxy properties are consistent with the passively-evolving
fundamental plane (FP) of early-type galaxies.  In this case the FP
estimate of the lens redshift is $z=1.0\pm 0.1$.

\section{Gravitational lens models}
\label{sec:models}

\subsection{Simple two-image models}
\label{subsec:models}

Constructing a model of the gravitational potential of the lens galaxy
is important for three reasons. First, if the model is physically
plausible, it corroborates our claim that \objC~is truly a lens.
Second, a lens model is required to predict the time delay between
lensed images as a function of the Hubble constant.  Third, if enough
constraints are available, the lens model may reveal interesting
aspects of the matter distribution of the lens galaxy, including dark
matter.

The present observations of \objC~provide five constraints on models
of the lens potential: the positions of A and B with respect to the
lens center, and the A/B radio flux density ratio.  The relative
separation of A and B is known precisely from the VLBA image.  Because
the position of radio component C is consistent with the optical lens
galaxy G, we assume that C marks the lens center, although we enlarge
the uncertainty to 10~mas to allow for the possibility of a small
displacement between the lens center and the radio component.  Our
measurements of the radio flux density ratio have a mean of 13.2, and
a scatter of 15\%.

The simplest plausible lens model is a singular isothermal sphere,
which produces two images on opposite sides of the lens, with a
magnification ratio equal to the ratio of distances from each image to
the lens center.  As stated earlier in \S\S~\ref{subsec:radio-discuss}
and~\ref{subsec:hst}, the present data are nearly consistent with this
model.  Component C lies only 7~mas from the line joining A and B, and
the ratio of distances is $13.8\pm 0.8$, in agreement with the flux
density ratio.

This confirms that gravitational lensing is a natural explanation for
the morphology of \objC.  Unfortunately there are not enough
constraints to explore less idealized lens models in detail.  A
singular isothermal ellipsoid (SIE), for example, has five parameters
(the position of the lens, the mass scale, the ellipticity and the
position angle), and is therefore uniquely determined by our five
constraints.  Table~\ref{tbl:model} gives the parameters of the SIE
model, as computed with lens modeling software written by
\citet{keeton01}.  The quoted uncertainties reflect the variations in
the parameters obtained by varying the position of the lens center and
the magnification ratio throughout their quoted error ranges.

The final quantity listed in Table~\ref{tbl:model}, labeled
$\Delta\tau$, is a dimensionless factor that may be converted into a
time delay $\Delta t$ by assuming a lens redshift and a cosmological
model as follows:
\begin{equation}
\Delta t = \frac{D_{l}D_{s}}{2c D_{ls}}(1+z_l) \Delta\tau.
\end{equation}
In this equation, $z_l$ is the lens redshift, and $D_{l}$, $D_{s}$ and
$D_{ls}$ are the angular-diameter distances to the lens, to the
source, and between the lens and source, respectively.

If we assume $z_{l} = 1.0\pm 0.1$, as predicted by the FP method (see
\S~\ref{subsec:spectrum}), then in a cosmology in which matter is
distributed smoothly with $\Omega_{\mathrm m} = 0.3$ and
$\Omega_\Lambda = 0.7$, the result is $h\Delta t = 118.5\pm 1.9$ days,
where $H_0 = 100h$~km~s$^{-1}$~Mpc$^{-1}$ as usual.  If instead
$\Omega_{\mathrm m} = 0.3$ and $\Omega_\Lambda = 0$, then $h\Delta t =
125.7\pm 2.0 $ days.  Image A is expected to lead image B.  The
uncertainties quoted here are internal to the SIE model only; the true
uncertainty in the time delay is much larger.

\subsection{The nature of component C}
\label{subsec:third}

We now return to the question of the origin of C, the third and
faintest radio component.  Its position relative to A and B agrees
with that of the galaxy G detected in the $I$-band HST image
(\S~\ref{subsec:hst}), so it possibly represents emission from a faint
AGN in the lens galaxy.  In this section we consider the implications
of another possibility---that C represents a third quasar image.

Although the SIS and SIE models only predict two images of the
background source, many lens models predict the existence of a third
image of the quasar located close to the lens center that is
demagnified by an amount depending on the central surface density of
the lens galaxy (see, e.g., \citealt{narasimha86}).  The existence or
absence of the ``odd image'' can therefore provide information on the
central structure of the lens galaxy, which would otherwise be very
difficult to obtain for a galaxy at $z\sim 1$.  The only lens for
which an odd image has been identified confidently is APM~08279+5255
\citep{ibata99,egami00}.  Many of the current observations of radio
lenses, which have typical dynamic ranges of 100--1000, show no
evidence for odd images; presumably the absence of the odd images is
due to the high central surface densities of the lens galaxies
\citep{wallington93,rusin01,keeton01b}.

There are two simple extensions of singular isothermal lens models
that produce three images. In the first category are models with a
finite core radius $r_c$.  In these models, the magnification of the
odd image is approximately $(\theta_c / \theta_E)^{2}$, where
$\theta_E$ is the Einstein ring radius of the lens galaxy
(approximately half the image separation), and $\theta_c = D_l r_c$,
the angle subtended by the core radius.  In our SIE model for \objC,
if C is an odd image then its magnification is $(111\pm 37)^{-1}$,
implying $(\theta_c / \theta_E) = 0.095\pm 0.016$.  For $z_l = 1$ this
corresponds to a core radius of $590\pm 100$ parsecs.  A core radius
this large would be interesting, because one would expect much smaller
(or zero) core radii based on the cusped distribution of starlight in
HST images of nearby galaxies \citep{faber97}, the absence of odd
images of most other lensed quasars
\citep{wallington93,rusin01,keeton01b}, and theoretical simulations of
cold-dark-matter haloes \citep{navarro96}.

For exactly these reasons, several authors have recently emphasized a
second category of lens models, which have a central density cusp
\citep[see, e.g.,][]{evans98,keeton01b,munoz01}.  If the cusp in the
three-dimensional mass density obeys the power law $\rho \sim
r^{-\gamma}$ for small $r$, then a third image is produced for $\gamma <
2$.  These models have even more trouble accomodating C as an odd
image.  For example, \citet{munoz01} applied a cusped lens model to
the 3-image lens APM~08279+5255, and found that, in order to match the
observational constraints, the cusp exponent had to be so shallow as
to approach the limit of models with a finite core radius.

In short, if C is a third quasar image, then the mass distribution of
the inner few hundred parsecs of the lens galaxy in \objC~is not as
centrally condensed as one would expect from studies of nearby
galaxies, studies of other lenses, and theoretical models of halo
formation.  It would therefore be interesting to test definitively
whether C is galaxy emission or a third quasar image.

Probably the simplest approach would be to measure the radio continuum
spectrum of C over as broad a wavelength range as possible.  If C is a
third image, its spectrum should match that of A and B.  There is one
indication in the present data that C may have a steeper spectrum than
that of A and B: the trend of increasing flux density ratio with radio
frequency that is depicted in the bottom panel of
Figure~\ref{fig:fluxes}.  The VLA data below 22.5~GHz do not have
sufficient angular resolution to distinguish B and C, so their flux
densities are lumped together in the models.  A steeper spectrum for C
would therefore cause the inferred flux density ratio to decrease with
frequency.  For example, if the true A/B flux density ratio is 14.7
(as implied by the April~2001 MERLIN data) and independent of
frequency, then in order to lower the flux density ratio to match the
1.4~GHz data, the required spectral index of C is $\alpha=-1.3$.

Unfortunately there are other reasons to expect that the measured flux
density ratio A/B is not completely independent of frequency, such as
variability (coupled with the time delay), and the possibility of a
large magnification gradient across a source with frequency-dependent
substructure.  It would be preferable to obtain multi-wavelength radio
images with sufficient angular resolution and sensitivity to resolve
all three components.

\section{Summary and future prospects}
\label{sec:summary}

Radio source \objC~is the fourth confirmed gravitational lens in our
survey of the southern sky for new lenses.  Here we briefly review the
evidence for lensing and the immediate prospects for using this lens
to obtain interesting astrophysical information.  Two compact,
flat-spectrum radio components separated by $1\farcs47$ are almost
certainly either a binary quasar or a pair of lensed images of a
single quasar.  The flux density ratio is within 15\% of 13.2 as
measured from 1.4~GHz to 43~GHz; this similarity of radio spectra
favors the lens hypothesis.  The lens galaxy is apparent in an optical
image with HST, and it is located exactly where one would expect,
based on the simplest plausible lens model.

The background source is a spectroscopically verified quasar at
redshift 3.42.  The optical magnitude, color, and effective radius of
the lens galaxy (together with the mass estimate provided by the image
separation) are consistent with a passively-evolving elliptical galaxy
at redshift unity.  There is a faint third radio component located at
or near the center of the lens galaxy, which is either faint emission
from the galaxy itself, or a third quasar image.  In the latter case,
the study of the faint component may be useful as a probe of the
central few hundred parsecs of a $z\sim 1$ galaxy.

The flat radio spectrum, and mild variability observed at 8.5 GHz,
make this system a candidate for flux monitoring, in order to use the
time delay between the quasar images to determine the Hubble constant.
The present data, however, do not provide nearly enough model
constraints for this method to be competitive with local
distance-scale methods.  One would need an additional, rich body of
constraints, possibly from infrared imaging of the quasar host galaxy
\citep[see, e.g.][]{kochanek01} or VLBI substructure \citep[see,
e.g.,][]{trotter00}.

\acknowledgments We are grateful to David Rusin, for a critical
reading of an early version of this report; Brian McLeod, for help
with HST image decomposition; Chuck Keeton, for making his lens
modeling code publicly available; Ed Turner, for arranging APO
spectroscopic observations; Phillip Helbig, for releasing archival VLA
data; Chris Fassnacht, for assistance with the VLBA; and Tom Muxlow
and Peter Thomasson, for their help with MERLIN. This research was
supported by the National Science Foundation under grants AST-9617028
and AST-9616866. J.N.W.\ thanks the Fannie and John Hertz foundation
for financial support.

\clearpage

\clearpage

\begin{deluxetable}{lcccccccccccc}
\rotate

\tabletypesize{\scriptsize}
\tablecaption{Radio observations of \objC\label{tbl:radio}}
\tablewidth{0pt}

\tablehead{
\colhead{Date}              &
\colhead{Observatory}       &
\colhead{Frequency}         &
\colhead{Bandwidth}         &
\colhead{Duration}          &
\colhead{Beam FWHM}         &
\multicolumn{3}{c}{Flux density}      &
\colhead{RMS noise}         &
\colhead{Abs.\ flux}        \\

\colhead{} &
\colhead{} &
\colhead{(GHz)} &
\colhead{(MHz)} &
\colhead{(min)} &
\colhead{(mas $\times$ mas, P.A.)} &
\colhead{A (mJy)} &
\colhead{B (mJy)} &
\colhead{C (mJy)} &
\colhead{(mJy/beam)} &
\colhead{uncertainty}
}

\startdata
1984 Dec 17 & VLA   & 4.860  & 100 & 1.3 & $496 \times 393$ ($-34\arcdeg$) & 206.01 & 17.06 & \nodata & 0.38 & 5\% \\
1994 Feb 24 & VLA   & 8.452  &  50 & 1.3 & $294 \times 198$ ($-33\arcdeg$) & 152.53 & 12.48 & \nodata & 0.26 & 3\% \\
1999 Jun 30 & VLA   & 8.440  & 100 & 1.5 & $743 \times 207$ ($54\arcdeg$)  & 138.90 & 12.21 & \nodata & 0.27 & 5\% \\
2000 Apr 05 & MERLIN& 4.994  & 15  & 110 & $95\times 42$ ($23\arcdeg$)     & 177.41 & 13.80 & $<2.9$  & 0.57 & 5\% \\
2000 Apr 29 & VLBA  & 4.987  & 32  & 54  & $3.4\times 1.4$ ($-2\arcdeg$)   & 166.21 & 11.93 & 0.50    & 0.14 & 5\% \\ 
2000 Sep 26 & ATCA  & 4.800  & 128 & 5   & $1385\times 9840$ ($0\arcdeg$)  & 216.62 & 16.48 & \nodata & 0.57 & 3\% \\
2000 Sep 26 & ATCA  & 6.080  & 128 & 10  & $1140\times2600$ ($0\arcdeg$)   & 186.44 & 14.29 & \nodata & 0.43 & 3\% \\
2000 Sep 26 & ATCA  & 8.640  & 128 & 15  & $791\times 2020$ ($0\arcdeg$)   & 164.18 & 12.56 & \nodata & 0.43 & 3\% \\
2000 Nov 11 & VLA   & 1.425  & 100 & 2.0 & $3430\times 1250$ ($53\arcdeg$) & 211.35 & 18.68 & \nodata & 0.30 & 5\% \\
2000 Nov 11 & VLA   &14.940  & 100 & 10  & $242 \times 120$ ($52\arcdeg$)  & 143.81 & 9.42  & $<1.5$  & 0.29 & 5\% \\
2000 Nov 11 & VLA   &22.460  & 100 & 10  & $184 \times 78$  ($54\arcdeg$)  & 126.48 & 8.36  & $<3.7$  & 0.74 &10\% \\
2000 Dec 24 & VLA   &42.640  & 100 & 12  & $62\times 38$ ($-44\arcdeg$)    & 83.42  & 5.43  & $<3.3$  & 0.66 &15\% \\
2001 Apr 13 & MERLIN& 4.994  & 15  & 525 & $88\times 44$ ($24\arcdeg$)     & 208.12 & 14.20 & 0.87    & 0.13 & 5\%
\enddata

\end{deluxetable}

\begin{deluxetable}{ccc}
\tabletypesize{\scriptsize}
\tablecaption{VLBA relative positions\label{tbl:radio-positions}}
\tablewidth{0pt}

\tablehead{
\colhead{Component} &
\colhead{$\Delta$R.A.~($x-x_{\mathrm A}$)} &
\colhead{$\Delta$Decl.~($y-y_{\mathrm B}$)} \\

\colhead{} &
\colhead{(mas)} &
\colhead{(mas)}
}

\startdata
A & \nodata & \nodata \\ 
B & $-1241.6\pm 0.1$ & $-784.6\pm 0.2$ \\
C & $-1160.6\pm 0.7$ & $-726.8\pm 1.7$
\enddata

\tablecomments{ The J2000 coordinates of component A are R.A.\
$=16^{\mathrm h}32^{\mathrm m}57\fs680$, Decl.\ $=-00\arcdeg 33\arcmin
21\farcs05$, within $0\farcs15$. }

\end{deluxetable}

\begin{deluxetable}{lccccc}
\tabletypesize{\scriptsize}
\tablecaption{Photometric model of \objC~based on HST/WFPC2 images\label{tbl:hst}}
\tablewidth{0pt}

\tablehead{
\colhead{Component} &
\colhead{$\Delta x$} &
\colhead{$\Delta y$} &
\colhead{$R_{\mbox{eff}}$} &
\colhead{$m_{\mbox{F555W}}$} &
\colhead{$m_{\mbox{F814W}}$} \\

\colhead{} &
\colhead{(mas)} &
\colhead{(mas)} &
\colhead{(arcsec)} &
\colhead{} &
\colhead{} \\
}

\startdata
A & 0              & 0              & \nodata       & $21.67\pm 0.06$ & $20.74\pm 0.06$ \\
B & $-1242.8\pm 3$ & $-790.9\pm 3$  & \nodata       & $24.97\pm 0.23$ & $24.4\pm 0.1$   \\
G & $-1161.6\pm 9$ & $-737.8\pm 9$  & $0.20\pm 0.08$& $\geq 25.5$     & $23.3\pm 0.2$
\enddata

\tablecomments{ Quoted uncertainties are based on statistical error
only, and are internal to the chosen model-fitting procedure.  In
particular the magnitude uncertainties do not include the overall
uncertainty in the zero-point or CTE correction (see
\S~\ref{subsec:hst}). }

\end{deluxetable}

\begin{deluxetable}{lcccc}
\tabletypesize{\scriptsize}
\tablecaption{Total optical magnitudes of \objC\label{tbl:optical}}
\tablewidth{0pt}

\tablehead{
\colhead{Filter} &
\colhead{Magnitude} \\
}

\startdata
$B$ & $22.85\pm 0.10$ \\
$V$ & $21.29\pm 0.05$ \\
$R$ & $20.99\pm 0.05$ \\
$I$ & $20.36\pm 0.10$
\enddata

\end{deluxetable}

\begin{deluxetable}{lcccc}
\tabletypesize{\scriptsize}
\tablecaption{Optical flux ratios\label{tbl:fluxratio}}
\tablewidth{0pt}

\tablehead{
\colhead{Filter} &
\colhead{A:(B+G)} \\
}

\startdata
$g'$ & $27 \pm 8$ \\
$r'$ & $24 \pm 3$ \\
$i'$ & $11.5 \pm 0.9$
\enddata

\end{deluxetable}

\begin{deluxetable}{ccccccc}
\tabletypesize{\scriptsize}
\tablecaption{SIE model parameters\label{tbl:model}}
\tablewidth{0pt}

\tablehead{
\colhead{Parameter} &
\colhead{Value} \\
}

\startdata
Einstein ring radius, $b$        & $0\farcs731\pm 0.015$ \\
Ellipticity, $\epsilon$          & $0.078^{+0.12}_{-0.02}$ \\
Position angle                   & $-4\arcdeg ^{+70}_{-28}$ \\
$x_{\mbox{source}} - x_{\mbox{A}}$   & $-0\farcs633\pm 0\farcs024$ \\
$y_{\mbox{source}} - y_{\mbox{A}}$   & $-0\farcs374\pm 0\farcs030$ \\
Magnification of A               & $2.11\pm 0.17$ \\
$\Delta\tau$                     & $0.932\pm 0.015$
\enddata

\tablecomments{ Quoted uncertainties reflect the range in parameter
values obtained by varying (separately) the lens position by 10~mas
and the magnification ratio by 15\%. }

\end{deluxetable}

\end{document}